\documentclass{appolb}%
\usepackage{epsfig}
\usepackage{rotating}
\usepackage{hyperref}
\usepackage{comment,color}
\usepackage{braket}
\usepackage{amssymb,amsmath,bm}
\usepackage{setspace,lipsum}
\usepackage{amsmath}
\usepackage{amsfonts}
\usepackage{amssymb}
\usepackage{graphicx}%
\providecommand{\U}[1]{\protect\rule{.1in}{.1in}}
\begin{document}

\title{Chiral anomaly role in $\pi_{1}(1600)\rightarrow\pi\eta^{\prime}$}
\author{Francesco Giacosa \address{Institute of Physics, Jan Kochanowski University, Kielce, Poland}}
\maketitle

\begin{abstract}
The ground-state (lightest) hybrid nonet with exotic quantum numbers
$J^{PC}=1^{-+}$ and the nonet of their chiral partners with $J^{PC}=1^{+-}$
build a homochiral multiplet involving left- and right-handed currents, which
under chiral transformation change just as (axial-)vector mesons. Masses and
interactions of hybrids can be obtained in the context of the extended Linear
Sigma Model. Here, we concentrate on the decays oh hybrids into two
pseudoscalar mesons, such as $\eta\pi$ and $\eta^{\prime}\pi$ modes. Indeed,
$\pi_{1}(1400)\rightarrow\pi\eta$ and $\pi_{1}(1600)\rightarrow\pi\eta
^{\prime}$ have been seen in experiments. Assuming that $\pi_{1}(1400)$ and
$\pi_{1}(1600)$ correspond to the same state $\pi_{1}^{hyb}$, we show that
these decays (and similar ones) follow from a chirally symmetric interaction
term that breaks explicitly the axial anomaly. In this respect, these decays
would be an additional manifestation of the axial (or chiral) anomaly in the
mesonic sector.

\end{abstract}


\section{Introduction}

Hybrids are unconventional mesons made of a quark-antiquark pair and a
constituent gluon. While they are predicted in many different approaches to
QCD, see e.g. Ref. \cite{meyer,lebed}, they could not be yet confirmed
experimentally (even if a substantial experimental effort is ongoing
\cite{compass,gluex}). Quite remarkably, in the PDG \cite{pdg} the two
enigmatic and very broad resonances $\pi_{1}(1400)$ and $\pi_{1}(1600)$ have
exotic quantum numbers $J^{PC}=1^{-+}$ (impossible for ordinary $\bar{q}q$
pairs). The state $\pi_{1}(1400)$ has been seen in the decay channels $\rho
\pi$ and $\eta\pi$, while the state $\pi_{1}(1600)$ in the channels
$b_{1}(1270)\pi,$ $f_{1}(1270)\pi,$ and $\eta^{^{\prime}}\pi.$ Recently, the
COMPASS experiment \cite{compass} confirmed the existence of $\pi_{1}(1600)$,
in particular by studying the decay $\pi_{1}(1600)\rightarrow\eta^{\prime}\pi
$. As we shall discuss below, this decay is one of the main subject of this work.

In the framework of lattice QCD exotic mesons with $J^{PC}=1^{-+}$ are the
lightest hybrid mesons and have a mass of about $1.6$ GeV \cite{michael,dudek}%
. Yet, only a single $\pi_{1}$ state is expected; it is then hard to
accommodated both $\pi_{1}(1400)$ and $\pi_{1}(1600)$. In Ref. \cite{rodas} it
was argued that the states $\pi_{1}(1400)$ and $\pi_{1}(1600)$ correspond to
the very same pole in the complex energy plane, and hence to a single state,
whose mass is close to $\pi_{1}(1600)$. Within this scenario, there is no
conflict with lattice and model predictions about the exotics and, as a
consequence, the unique $\pi_{1}^{hyb}$ resonance -to be identified with
$\pi_{1}(1600)$- decays both into $\eta\pi$ and $\eta^{^{\prime}}\pi.$

The first immediate question concerning hybrids is the following: if the
$\pi_{1}(1600)$ is the lightest hybrid meson with isospin 1, where are the
other members of the multiplet? Namely, hybrids form nonets just as regular
states. As a consequence, one expects two states with $I=0,$ denoted as
$\eta_{1,N}^{hyb}$ (predominantly nonstrange) and $\eta_{1,S}^{hyb}$
(predominantly strange)) as well as four states with $I=1/2$ collectively
denoted as $K_{1}^{hyb}$. Their mass should be also  close to $1.6$ GeV.

Moreover, chiral symmetry also implies that chiral partners of this low-lying
hybrid nonet should exist. The corresponding quantum numbers are
$J^{PC}=1^{+-}$(just as pseudovector mesons, hence we deal with cryptoexotic
states). Their mass should be at about 2 GeV (or heavier).

An attempt to answer these questions was recently presented in Ref.
\cite{hybrid}. The ground-state hybrids and their chiral partners form a
chiral multiplet which can be coupled to ordinary $\bar{q}q$ states as
(pseudo)scalar and (axial-)vector mesons. This is achieved in the framework of
a well established chiral model of QCD, the so-called extended Linear Sigma
Model (eLSM). In this model, symmetries of QCD (and their violations) are
implemented at the level of composite hadrons. As shown in Ref. \cite{dick},
a  fit using masses and decays of various mesons up to 1.8 GeV shows a good
agreement with experimental data. In addition, also baryonic d.o.f.
\cite{gallas,olbrich}, various glueballs \cite{glueballselsm}, extensions to
excited states \cite{excited} as well as studies at nonzero temperature and
densities \cite{kovacs} have been developed.

It seems then natural to use the eLSM to evaluate decays of hybrids,
especially in order to find out which decays are favoured and which ones are
suppressed. Moreover, various ratios among decays represent clear predictions
of the approach. As shown in\ Ref. \cite{hybrid} one of the dominant decays is
found to be $\pi_{1}(1600)\rightarrow b_{1}(1230)\pi$, in agreement with
lattice \cite{michael} and with other model predictions \cite{page}.

In this work, we concentrate on the decay into $\pi_{1}(1600)$ into $\eta\pi$
and $\eta^{\prime}\pi$. Quite interestingly, this decay turns out to be
possible only through an interaction term that breaks the axial symmetry. The
importance of this so-called chiral (or axial) anomaly is well appreciated for
the masses and mixing of the mesons $\eta$and $\eta^{\prime}$\cite{thooft}.
Yet, this anomaly can affect also other parts of the hadronic spectrum, as
recently discussed in\ Refs. \cite{olbrich,pisarski}. Along this line, we show
that the axial anomaly can be also responsible for the decays of hybrids into
$\eta\pi$ and $\eta^{\prime}\pi$.

\section{Fields and model}

Here, we briefly review the model presented in Ref. \cite{hybrid} and its results.

First, we recall the (pseudo)scalar sector. The $3\times3$ matrix $P$ contains
the light pseudoscalar nonet \{$\pi$, $K,\eta,\eta^{\prime}$\} with quantum
numbers $J^{PC}=0^{-+}$\cite{pdg}. At a fundamental level, it is made of
quark-antiquark elements given by $P_{ij}=2^{-1/2}\bar{q}_{j}i\gamma^{5}q_{i}%
$with $i,j=u,d,s$. The matrix $S$, whose $\bar{q}q$ elements are the scalar
currents $S_{ij}=2^{-1/2}\bar{q}_{j}q_{i}$, contains the scalar fields
\{$a_{0}(1450),$ $K_{0}^{\ast}(1430),$ $\sigma_{N}\approx f_{0}(1370),$
$\sigma_{S}\approx f_{0}(1710)$\} with $J^{PC}=0^{++}$. The scalar and
pseudoscalar matrices are combined into the matrix $\Phi=S+iP$, which under
chiral transformations $U_{L}(3)\times U_{R}(3)$ changes as $\Phi\rightarrow
U_{L}\Phi U_{R}^{\dagger}$ ($U_{L}$and $U_{R}$ being $3\times3$ unitary
matrices). Under parity: $\Phi\rightarrow\Phi^{\dagger}$ and under charge
conjugation (denoted as $C$): $\Phi\rightarrow\Phi^{t}$.

Next, we consider pseudovector and excited vector states. They follow from the
(pseudo)scalar currents upon introducing a derivative in between the quarks.
The nonet $B^{\mu}$ with elements $B_{ij}^{\mu}=2^{-1/2}\bar{q}_{j}\gamma
^{5}\partial^{\mu}q_{i}$ has quantum numbers $J^{PC}=1^{+-}$and describes the
fields \{$b_{1}(1230),$ $K_{1}(1270)$/$K_{1}(1400),$ $h_{1}(1170),$
$h_{1}(1380)$\}, see Ref. \cite{hybrid} for details. This nonet, together with
the nonet of orbitally excited vector mesons $V_{E,ij}^{\mu}=2^{-1/2}\bar
{q}_{j}i\partial^{\mu}q_{i}$ involving the resonances \{$\rho(1700)$,
$K^{\ast}(1680),$$\omega(1650)$, $\phi(1930?)$\},builds the chiral multiplet,
$\tilde{\Phi}^{\mu}=V_{E}^{\mu}-iB^{\mu}$, which transforms just as
(pseudo)scalar fields under chiral transformations: $\tilde{\Phi}^{\mu
}\rightarrow U_{L}\tilde{\Phi}^{\mu}U_{R}^{\dagger}$. This is a consequence of
the fact that the derivative does not modify the chiral properties; in
general, multiplets transforming in this way are called heterochiral
\cite{pisarski}. Moreover: $\tilde{\Phi}^{\mu}\rightarrow\tilde{\Phi}_{\mu
}^{\dag}$ under parity and $\tilde{\Phi}^{\mu}\rightarrow-\tilde{\Phi}^{t\mu}$
under $C.$

We turn to (axial-)vector states. The matrix $V^{\mu},$with $V_{ij}^{\mu
}=2^{-1/2}\bar{q}_{j}\gamma^{\mu}q_{i}$, carries the vector mesons
\{$\rho(770),$ $K^{\ast}(892),$ $\omega(782),$ $\phi(1020)$\} with
$J^{PC}=1^{--}$. Analogously, the matrix $A^{\mu}$, with $A_{ij}^{\mu
}=2^{-1/2}\bar{q}_{j}\gamma^{5}\gamma^{\mu}q_{i},$contains the axial-vector
mesons \{$a_{1}(1230),$ $K_{1}(1270)$/$K_{1}(1400),$ $f_{1}(1285),$
$f_{1}(1420)$\} with $J^{PC}=1^{++}$. These matrices are combined into the
right- and left-handed combinations $R^{\mu}=V^{\mu}-A^{\mu}$ and $L^{\mu
}=V^{\mu}+A^{\mu}$ which under chiral transformation behave as $R^{\mu
}\rightarrow U_{R}R^{\mu}U_{R}^{\dagger}$ and $L^{\mu}\rightarrow U_{L}L^{\mu
}U_{L}^{\dagger}$, thus in an utterly different way w.r.t. to (pseudo)scalars.
We refer to them as an homochiral multiplet \cite{pisarski}. Under parity:
$R^{\mu}\rightarrow L_{\mu}$and $L^{\mu}\rightarrow R_{\mu}$; under $C$:
$R^{\mu}\rightarrow-L^{\mu t}$ and $L^{\mu}\rightarrow-R^{\mu t}.$

Next, one builds objects analogous to the (axial-)vector fields in the hybrid
sector, upon including the gluon field. To this end, we consider the objects
\begin{equation}
\Pi^{hyb,\mu}\equiv2^{-1/2}\bar{q}_{j}G^{\mu\nu}\gamma_{\nu}q_{i}\text{ and
}B_{ij}^{hyb,\mu}=2^{-1/2}\bar{q}_{j}G^{\mu\nu}\gamma_{\nu}\gamma^{5}q_{i}%
\end{equation}
which -besides the standard quark-antiquark pair- involve also explicitly the
gluon field strength tensor $G^{\mu\nu}$, being responsible for the switch of
the $C$-parity. As a consequence, the nonet $\Pi^{hyb,\mu}$ has exotic quantum
numbers $J^{PC}=1^{-+}$; the corresponding nonet is denoted as \{$\pi
_{1}(1600),$ $K_{1}(?),$ $\eta_{1}(?),$ $\eta_{1}(?)$\}$,$ where -at present-
only the isovector member can be assigned to a physical resonance. The nonet
$B^{hyb,\mu}$ with $J^{PC}=1^{+-}$ contains \{$b_{1}(?),$ $K_{1,B}(?),$
$h_{1}(?),$ $h_{1}(?)$\}, for which there are not yet experimental candidates.
These two nonets are grouped into the right-handed and left-handed currents
$R^{hyb,\mu}=\Pi^{hyb,\mu}-B^{hyb,\mu}$ and $L^{hyb,\mu}=\Pi^{hyb,\mu
}+B^{hyb,\mu}$, which transform as $R^{hyb,\mu}\rightarrow U_{R}R^{hyb,\mu
}U_{R}^{\dagger}$ and $L^{hyb,\mu}\rightarrow U_{L}L^{hyb,\mu}U_{L}^{\dagger}$
(just as (axial-)vectors). Moreover: $R^{hyb,\mu}\rightarrow L_{\mu}^{hyb}$
and $L^{hyb,\mu}\rightarrow R_{\mu}^{hyb}$ under parity and $R^{hyb,\mu
}\rightarrow L_{\mu}^{hyb,t}$ and $L^{hyb,\mu}\rightarrow R_{\mu}^{hyb,t}$
under $C.$ (For a list of other possible multiplets together with their
homo/heterochirality, see the classification of Ref. \cite{pisarski}.)

In Ref. \cite{hybrid} a Lagrangian that couples the hybrid multiplet to
conventional mesons is presented. Both masses and decays of hybrids can be
described. For instance, a term proportional to $\mathrm{Tr}\left(  L_{\mu
}^{hyb}\Phi R^{hyb,\mu}\Phi^{\dagger}\right)  $ is present. It is invariant
under $U_{L}(3)\times U_{R}(3)$ as well as parity and $C$, as one can verify
by using the transformations above; it is important since it generates the
mass difference between the hybrids with $J^{PC}=1^{-+}$ and with
$J^{PC}=1^{-+}.$ Namely, one finds $m_{b_{1}^{hyb}}^{2}-m_{\pi_{1}^{hyb}}%
^{2}\propto\phi_{N}^{2}\,,$ where $\phi_{N}$ is the chiral condensate emerging
from the spontaneous symmetry breaking of chiral symmetry. Thus, just as for
ordinary mesons, also for hybrids the mass splitting between chiral partners
is generated by the chiral condensate. Finally, one gets $m_{\pi_{1}^{hyb}%
}\simeq m_{\eta_{1,N}^{hyb}}\simeq1660$ MeV, $m_{K_{1}^{hyb}}\simeq1707$ MeV,
and $m_{\eta_{1,S}^{hyb}}\simeq1751$ MeV for the lighter $1^{-+}$ nonet, and
$m_{h_{1,N,B}^{hyb}}\simeq m_{b_{1}^{hyb}}\simeq2000$ MeV, $m_{K_{1,B}^{hyb}%
}\simeq2063$ MeV, $m_{h_{1,S}^{hyb}}\simeq2126$ MeV for the heavier $1^{+-}$ nonet.

For what concerns decays, four terms can be built \cite{hybrid}. The first two
terms fulfills chiral and dilatation invariance and are expected to be
dominant: the first terms is responsible for e.g. $\pi_{1}\rightarrow
b_{1}(1230)\pi$ , the second for e.g. $b_{1}^{hyb}\rightarrow\pi\pi\eta$ and
$\pi_{1}^{hyb}\rightarrow\pi\rho\eta$. The third term breaks dilatation
invariance and gives rise to $\pi_{1}^{hyb}\rightarrow\rho\pi$ and $\pi
_{1}^{hyb}\rightarrow K^{\ast}K$. Finally, the fourth one generates $\pi
_{1}^{hyb}\rightarrow\eta\pi$ and $\pi_{1}^{hyb}\rightarrow\eta^{\prime}%
\pi.\ $This decay, even if small, is important since the final state is easily
measurable.\ We study it studied in more details in the next section.

\section{Anomalous decays of hybrids}

The interaction term that describes $\pi_{1}^{hyb}\rightarrow\eta\pi$ and
$\pi_{1}^{hyb}\rightarrow\eta^{\prime}\pi$ cannot be constructed by a term
that fulfills $U_{L}(3)\times U_{R}(3).$ Yet, one can construct it by breaking
the $U_{A}(1)$ symmetry, and thus implementing the chiral anomaly. The
corresponding Lagrangian
\begin{equation}
\mathcal{L}_{eLSM}^{\text{ hybrid-anomaly}}=\beta_{A}^{hyb}(\det\Phi-\det
\Phi^{\dag})\mathrm{Tr}(L_{\mu}^{hyb}\left(  \partial^{\mu}\Phi\Phi^{\dag
}-\Phi\partial^{\mu}\Phi^{\dag}\right)  -R_{\mu}^{hyb}(h.c.))
\end{equation}
fulfills $SU_{L}(3)\times SU_{R}(3)$ and also parity and $C.$ Since it
involves the determinant, it explicitly breaks $U_{A}(1)$. Using $\det
\Phi-\det\Phi^{\dag}\propto\eta_{0}+...$ ($\eta_{0}$ is the flavor-singlet
combination) \cite{olbrich}, one obtains $\mathcal{L}_{eLSM}^{\text{
hybrid-anomaly}}\propto\eta_{0}\mathrm{Tr}(\Pi_{\mu}^{hyb}\partial^{\mu
}P)+...$ . Then, decays of the type $\pi_{1}^{hyb}\rightarrow\eta\pi$ and
$\pi_{1}^{hyb}\rightarrow\eta^{\prime}\pi$ emerge. Quite importantly, one can
predict the ratio $\Gamma_{\pi_{1}^{hyb}\rightarrow\eta^{\prime}\pi}%
/\Gamma_{\pi_{1}^{hyb}\rightarrow\eta\pi}\simeq12.7$, showing that the
$\eta^{\prime}\pi$ channel is favoured. In Eq. (\ref{simtable}) we present the
results of the ratios for all the nonet members.%
\begin{align}
\frac{\Gamma_{\pi_{1}^{hyb}\rightarrow\pi\eta^{\prime}}}{\Gamma_{\pi_{1}%
^{hyb}\rightarrow\pi\eta}} &  =12.7\text{ , }\frac{\Gamma_{K_{1}%
^{hyb}\rightarrow K\eta}}{\Gamma_{\pi_{1}^{hyb}\rightarrow\pi\eta}}=0.69\text{
, }\frac{\Gamma_{K_{1}^{hyb}\rightarrow K\eta^{\prime}}}{\Gamma_{\pi_{1}%
^{hyb}\rightarrow\pi\eta}}=5.3\text{ ,}\nonumber\\
\frac{\Gamma_{\eta_{1,N}^{hyb}\rightarrow\eta\eta^{\prime}}}{\Gamma_{\pi
_{1}^{hyb}\rightarrow\pi\eta}} &  =2.2\text{ , }\frac{\Gamma_{\eta_{1,S}%
^{hyb}\rightarrow\eta\eta^{\prime}}}{\Gamma_{\pi_{1}^{hyb}\rightarrow\pi\eta}%
}=1.57\text{.}\label{simtable}%
\end{align}

\section{Conclusions}

The identification of $\pi_{1}(1600)$ as an exotic hybrid state implies that a
full nonet of hybrids as well as a nonet of chiral partners should exist. We
have investigated the chiral properties of hybrids and coupled them to
ordinary $\bar{q}q$ states in order to evaluate masses and decays. In
particular, we have concentrated on the decays into two pseudoscalar mesons,
such as $\pi_{1}^{hyb}\rightarrow\eta^{\prime}\pi$ and $\pi_{1}^{hyb}%
\rightarrow\eta\pi.$ These decays are a consequence of an interaction term
that breaks axial symmetry and thus may represent an interesting additional
manifestation of chiral anomaly in the mesonic sector. The ratios reported
in\ Eq. (\ref{simtable}) can be verified/falsified in ongoing and future experiments.

\bigskip

\textbf{Acknowledgments}: The author thanks W.\ Eshraim, D. Parganlija, and C.
Fischer for cooperation leading to Ref. \cite{hybrid}. Moreover, the author
acknowledges support from the Polish National Science Centre NCN through the
OPUS projects no. 2019/33/B/ST2/00613 and no. 2018/29/B/ST2/02576.

\end{document}